\documentclass[12pt]{article}
\usepackage{amsmath,amssymb}
\def\be{\begin{equation}}
\def\ee{\end{equation}}
\def\bea{\begin{eqnarray}}
\def\eea{\end{eqnarray}}

\def\kslash{\slash \hspace{-0.2cm} k}

\def\exp{e^{\bar\zeta {\slash \hspace{-0.15cm} k} \lambda}}
\begin{document}
\begin{titlepage}

\setcounter{page}{0}


\begin{flushright}
KEK-TH-989 \\
OIQP-04-06
\end{flushright}

\vskip 5mm
\begin{center}
{\Large\bf Wilson Loops and Vertex Operators in Matrix Model \\}
\vskip 15mm

{\large
 Satoshi Iso$^{*}$\footnote{\tt satoshi.iso@kek.jp},
Hidenori Terachi$^{*}$\footnote{\tt terachi@post.kek.jp}
and Hiroshi Umetsu$^\dagger$\footnote{\tt hiroshi\_umetsu@pref.okayama.jp}
} 
\vskip 13mm
$^*${\it Institute of Particle and Nuclear Studies \\
High Energy Accelerator Research Organization (KEK) \\
Oho 1-1, Tsukuba, Ibaraki 305-0801, Japan \\
and \\
Department of Particle and Nuclear Physics, \\
the Graduate University  for Advanced Studies (Sokendai)\\
Oho 1-1, Tsukuba, Ibaraki 305-0801, Japan} 

\vspace{5mm}

$^\dagger${\it Okayama Institute for Quantum Physics \\
 1-9-1 Kyoyama, Okayama City, Okayama 700-0015, Japan} \\
\end{center}

\vskip 30mm
\centerline{{\bf{Abstract}}}
\vskip 3mm
We systematically construct wave functions and vertex operators 
in the type IIB (IKKT) matrix model by expanding a supersymmetric 
Wilson loop operator.
They form a massless multiplet of  the ${\cal N}=2$ type IIB supergravity
and automatically satisfy conservation laws.

\end{titlepage}

\newpage

\setcounter{footnote}{0}
\setcounter{equation}{0}

\section{Introduction}
Type IIB (IKKT) matrix model was proposed as a nonperturbative 
formulation of superstrings \cite{IKKT} and has been expected to be equivalent 
to the type IIB superstring.
The Schwinger-Dyson equation of the Wilson lines are shown to describe
the string field equation of motion of type IIB superstring 
in the light cone gauge\cite{FKKT}.
Although there are still many issues to be resolved, the model has 
an advantage to other formulations of superstrings that we can discuss 
dynamics of space-time more directly\cite{AIKKT}.
The action of the model is given by
\be
S_{\mathrm{IKKT}} = -{1 \over 4}\mathrm{tr}~[A_\mu,A_\nu]^2 
- {1 \over 2} \mathrm{tr}~\bar{\psi}\Gamma^\mu [A_\mu,\psi],
\label{IKKT}
\ee
where $A^{\mu}$ ($\mu =0, \cdots,9$) and ten-dimensional Majorana-Weyl 
fermion $\psi$ are $N \times N$ bosonic and fermionic hermitian matrices. 
The action was originally derived from the Schild action for the 
type IIB superstring by regularizing the world sheet coordinates by matrices.
It is interesting that 
the same action describes the effective action for $N$ D-instantons\cite{Witten}.
This suggests a possibility that D-instantons (D(-1)) can be considered 
as fundamental objects to generate both of the space-time and 
the dynamical fields (or strings) on the space-time.
The bosonic matrices represent noncommutative coordinates of D(-1)'s and 
the distribution of eigenvalues of $A_\mu$ is interpreted to form space-time.

If we take the above interpretation that the space-time is constructed 
by distribution of D-instantons, how can we interpret the $SO(9,1)$ 
rotational symmetry of the matrix model action?
This symmetry can be interpreted in the sense of mean field.
Namely we can consider that the system of $N$ D-instantons are embedded in
larger size 
$(N+M)\times (N+M)$ matrices as 
\be
\left(
\begin{array}{cc}
N D(-1) & \\
& M D(-1) \mbox{ as background for } N D(-1) 
\end{array} \right),
\label{NM}
\ee
and consider the action (\ref{IKKT}) as an effective action 
in the background where the rest, $M$ eigenvalues, distribute uniformly 
in 10 dimensions.
If the $M$ eigenvalues distribute inhomogeneously,
we may expect that the effective action for $N$ D-instantons is modified
so that they live in a curved space-time.
This is analogous to a thermodynamic system. In a canonical ensemble,
a subsystem in a heat bath is characterized by several thermodynamic 
quantities like temperature. Similarly a subsystem 
of $N$ D-instantons in a ``matrix bath'' can be characterized by several 
thermodynamic quantities. 

Since the matrix model has the ${\cal N}=2$ type IIB supersymmetry
\begin{eqnarray}
\left\{
\begin{array}{rl}
\delta A_\mu =& i \bar\varepsilon \Gamma_\mu \psi, \\
\delta \psi =& -{i \over 2} [A_\mu , A_\nu] \Gamma^{\mu\nu} \epsilon + \epsilon' 1_{N},
\end{array}
\right. \label{susy}
\end{eqnarray}
we expect that the configuration of the $M$ D-instantons 
can describe condensation of massless fields of the type IIB supergravity 
and the thermodynamic quantities of the matrix bath are characterized by the values
of the condensations.

In order to  discuss which type of configurations for $M$ D-instantons
correspond to the condensation of massless type IIB supergravity multiplet,
we consider the supersymmetry transformations (\ref{susy}) in the system
of $N+M$ D-instantons (\ref{NM}). 
In particular, we consider in this paper the simplest case that the background 
is represented by one D-instanton (namely $M=1$).
This simplification can be considered as a mean field approximation 
that the configuration of $M$ D-instantons is represented by a  
mean field described by a single D-instanton.
We call this extra D-instanton a {\bf mean field D-instanton}.
This kind of idea  was first discussed by Yoneya in \cite{Yoneya}. 
We hence embed $N \times N$ matrices into $(N+1) \times (N+1)$ matrices
as \bea
A_\mu^\prime =
\left(
\begin{array}{cc}
A_\mu & a_\mu \\
a_\mu^\dagger & y_\mu
\end{array}
\right), \qquad
\psi^\prime =
\left(
\begin{array}{cc}
\psi & \varphi \\
\varphi^\dagger & \xi
\end{array}
\right).
 \label{matrix}
\eea
Here we use  $A_\mu^{\prime}$, $\psi^{\prime}$ for $(N+1)\times (N+1)$
matrices and $A_\mu$, $\psi$ for $N \times N$ parts of the matrices.
$(y, \xi)$ is the coordinate of the mean field D-instanton
and its configuration (or the wave function) $f(y,\xi)$
specifies a certain background of the massless type IIB supergravity 
multiplet. The supersymmetry transformation  (\ref{susy}) 
for ($A_\mu^{\prime}$, $\psi^{\prime}$) can be rewritten in components as
\bea
\left\{
\begin{array}{rl}
\delta A_\mu &= i \bar\epsilon \Gamma_\mu \psi, \\
\delta y_\mu &= i \bar\epsilon \Gamma_\mu \xi, \\
\delta a_\mu &= i \bar\epsilon \Gamma_\mu \varphi,
\end{array}
\right.
\eea
and 
\bea
\left\{
\begin{array}{rl}
\delta\psi &= -{i \over 2} (F_{\mu\nu} + a_\mu a_\nu^\dagger - a_\nu a_\mu^\dagger)
\Gamma^{\mu\nu} \epsilon + \epsilon^\prime 1_N,
\\
\delta\xi &= -{i \over 2}(a_\mu^\dagger a_\nu - a_\nu^\dagger a_\mu )\Gamma^{\mu\nu}
\epsilon +\epsilon^\prime, 
\\
\delta\varphi &= -{i \over 2} \left\{ (A_\mu-y_\mu) a_\nu - (A_\nu-y_\nu) a_\mu \right\}
\Gamma^{\mu\nu}\epsilon,
\end{array}
\right.
\eea
where $F_{\mu  \nu}=[A_\mu, A_\nu]$.
We can obtain an effective action for the diagonal blocks by integrating
the off-diagonal parts $a_\mu$, $\phi$. 
In the leading order of the perturbation, we can neglect 
terms depending on the off-diagonal fields and the susy transformations
are given by
\bea
\left\{
\begin{array}{rl}
&\delta A_\mu = i \bar\epsilon \Gamma_\mu \psi, \\
&\delta \psi =-{i \over 2 } F_{\mu\nu} \Gamma^{\mu\nu} \epsilon+
 \epsilon^\prime,  \\
\end{array}
\right. \label{susy2}
\eea
\bea
\left\{
\begin{array}{rl}
\delta y_\mu &= i \bar\epsilon \Gamma_\mu \xi, \\
\delta \xi &= \epsilon^\prime. \\
\end{array}
\right. \label{susy3}
\eea
The first transformations for $N$ D-instantons are the same as 
the original susy transformations, eq. (\ref{susy}).
The second ones are ${\cal N}=2$ supersymmetry 
transformations for the single mean field D-instanton.
The generators of the former are given by
\bea
\bar\epsilon Q_1 
&=& i(\bar\epsilon \Gamma_\mu \psi){\delta \over \delta A_\mu}
- {i\over2}F_{\mu\nu}\Gamma^{\mu\nu}\epsilon {\delta \over \delta\psi},
\\
\bar\epsilon Q_2 &=& \epsilon {\delta \over\delta\psi},
\eea
while those of the latter are given by 
\bea
\bar{\epsilon}  q_1 &=& i \bar{\epsilon} \Gamma_\mu \xi {\partial \over \partial y_\mu} ~, \\
\bar{\epsilon}' q_2  &=& \epsilon' {\partial \over \partial \xi}. 
\eea
In order to obtain the correct wave functions $f(y,\xi)$ corresponding to the
massless supergravity multiplet, we need to obtain the multiplet of wave functions
that transform correctly under the supersymmetry transformation (\ref{susy3}).

When the supersymmetry transformations (\ref{susy3}) 
act on  wave functions of the form $e^{-ik \cdot y} f(\xi)$,
they  become
\bea
\bar\epsilon q_1 f(\xi) e^{-ik \cdot y}  &=& 
(\bar\epsilon  {\slash \hspace{-0.2cm} k} \xi )\ f(\xi)
e^{-ik \cdot y},  \nonumber \\
  \bar\epsilon' q_2 f(\xi) e^{-ik \cdot y}  &=& 
 \epsilon' {\partial \over \partial \xi} \ f(\xi)
e^{-ik \cdot y}. 
\label{susy4}
\eea

In this paper, as a first step toward discussing dynamics of condensation, 
we derive a multiplet of wave functions
for the single (mean field) D-instanton and corresponding
vertex operators for the $N$ D-instantons.
Such vertex operators were partly obtained by Kitazawa\cite{Kitazawa:2002vh} 
for the type IIB matrix model
by using the supersymmetry transformations.
We give a more systematic derivation of the vertex operators
by expanding supersymmetric Wilson loop operators.
They automatically form a supersymmetry multiplet and satisfy 
conservation laws.
In the context of M(atrix) theory for D-particles
or membrane theory, such vertex operators corresponding to the 
supergravity modes were also constructed by one-loop calculations
in fermionic backgrounds\cite{Taylor:1998tv} or 
by using supersymmetry transformations of the Wilson 
loop operator\cite{Dasgupta:2000df}.
Vertex operators for matrix strings were also constructed in 
\cite{matrixstring} by duality transformations of  the above.

The content of the paper is as follows.
In section 2, we construct a supersymmetric Wilson loop operator 
which is invariant under simultaneous  supersymmetry transformations 
of $N$ D-instantons and the remaining single D-instanton coordinate.
In section 3, by using the supersymmetry transformations,
we construct a multiplet of wave functions for the single D-instanton.
In section 4, we then expand the supersymmetric Wilson loop in terms of 
the wave functions derived in section 3 and obtain a multiplet of 
vertex operators for $N$ D-instantons. 
Section 5 is devoted to further discussions concerning dynamics of 
condensations. In appendix, we summarize our notations and useful 
identities.

\setcounter{equation}{0}
\section{Supersymmetric Wilson Loop}

In order to construct wave functions $f_A(\xi)  e^{-ik \cdot y}$ and
vertex operators $V_A(A^\mu, \psi;k)$ that transform 
covariantly under supersymmetries (\ref{susy3}) and (\ref{susy2}) respectively
($A$ denotes a field of a massless ${\cal N}=2$ supergravity multiplet),
we first consider a supersymmetric Wilson loop operator first introduced 
in \cite{Hamada} for the IIB matrix model;
\be
w(C) = \mathrm{tr} \prod_{j=1} e^{\bar\lambda_j Q_1} e^{-i \epsilon k_j^\mu A_\mu} 
e^{-\bar\lambda_j Q_1}.
\ee
Since we are interested in the massless multiplet, 
we here consider the following simplest Wilson loop operator
\be
\omega(\lambda,k)= e^{\bar\lambda Q_1}~ \mathrm{tr} e^{ik \cdot A} ~e^{- \bar\lambda Q_1}.
\label{omega}
\ee 
We will then show that by expanding the operator $\omega(\lambda,k)$ 
we can obtain a set of wave functions and vertex operators.
Hereafter, we assume that the $N\times N$ matrices $A_\mu$ and
$\psi$ satisfy the equations of motion,
\begin{eqnarray}
 && \label{eom-boson}
  \left[A^\nu, \left[A_\mu, A_\nu\right]\right]
  -\frac{1}{2}\left(\Gamma_0 \Gamma_\mu\right)_{\alpha\beta}
  \left\{\psi_\alpha, \psi_\beta\right\}=0, \\
 && \label{eom-fermion}
  \Gamma^\mu\left[A_\mu, \psi\right]=0.
\end{eqnarray}

First we show that $\omega(\lambda,k)$ is invariant 
under simultaneous supersymmetry transformations for
$N \times N$ matrices $A^\mu, \psi$ and the parameters $(\lambda, k)$.
When we act supersymmetry transformation $e^{\bar\epsilon Q_1}$
on $\omega(\lambda,k)$, it becomes 
\bea
e^{\bar\epsilon Q_1} \omega(\lambda,k) e^{-\bar\epsilon Q_1}
&=&
e^{\bar\epsilon Q_1} e^{\bar\lambda Q_1}~ \mathrm{tr} e^{ik \cdot A} ~e^{- \bar\lambda Q_1}e^{-\bar\epsilon Q_1} \nonumber\\
&=&
e^{(A^\mu \bar\epsilon \Gamma_\mu \lambda)G} e^{(\bar\epsilon + \bar\lambda)Q_1}
~\mathrm{tr} e^{ik \cdot A}~ 
e^{-(\bar\lambda + \bar\epsilon)Q_1} e^{(A^\nu \bar\epsilon \Gamma_\nu \lambda)G}
\nonumber\\
&=&
\omega (\epsilon + \lambda, k). \label{invariance1}
\eea 
Here $G$ is the generator of $U(N)$ transformation and we have used 
the commutation relation 
\bea
\label{susycomm11}
[\bar\epsilon_1 Q_1, \bar\epsilon_2 Q_1]
&=& 2 A^\mu \bar\epsilon_1 
\Gamma_\mu \epsilon_2 \ G 
\nonumber \\
&&+\left(
-{7\over8}(\bar\epsilon_1\Gamma^\mu \epsilon_2)\Gamma_\mu
+
\frac{1}{16\cdot 5!}(\bar\epsilon_1\Gamma^{\mu_1\cdots \mu_5}\epsilon_2)
\Gamma_{\mu_1\cdots \mu_5}
\right)
\Gamma_\lambda[A^\lambda, \psi]\frac{\delta}{\delta\psi}.
\eea
The second term on the right hand side vanishes due to the equation of
motion (\ref{eom-fermion}). 
Similarly for the other supersymmetry transformation 
$e^{\bar\epsilon Q_2}$, the Wilson loop operator 
transforms as  
\bea
e^{\bar\epsilon Q_2} \omega(\lambda,k) e^{-\bar\epsilon Q_2}
&=&
e^{\bar\epsilon Q_2} e^{\bar\lambda Q_1}~ \mathrm{tr} e^{ik \cdot A} ~e^{- \bar\lambda Q_1}e^{-\bar\epsilon Q_2} \nonumber\\
&=&
e^{\bar\lambda Q_1}e^{\bar\epsilon Q_2} 
e^{i(\bar\lambda \Gamma_\mu \epsilon){\delta \over \delta A_\mu}}
\mathrm{tr}e^{ik \cdot A}
e^{-i(\bar\lambda \Gamma_\mu \epsilon){\delta \over \delta A_\mu}}
e^{-\bar\epsilon Q_2}e^{-\bar\lambda Q_1} \nonumber \\
&=&
e^{-(\bar\lambda {\slash \hspace{-0.15cm} k} \epsilon)} \omega(\lambda,k), 
\label{invariance2}
\eea
where we have used the commutation relation
\be
[\bar\epsilon_1 Q_1, \bar\epsilon_2 Q_2] = -i \ (\bar\epsilon_1 \Gamma^\mu
\epsilon_2)  \frac{\partial}{ \partial A^\mu}. 
\ee

{}From (\ref{invariance1}) and (\ref{invariance2}), the following two
relations for the supersymmetric Wilson loop operator are obtained;
\begin{eqnarray}
 && \label{susywilsoninv1}
  [ \bar\epsilon Q_1, \omega(\lambda,k) ]
  -\epsilon{\partial \over \partial \lambda}\omega(\lambda,k) = 0,  \\
 && \label{susywilsoninv2}
  [ \bar\epsilon Q_2, \omega(\lambda,k) ]
  +(\bar\lambda \kslash \epsilon)\omega(\lambda,k)
  = 0.
\end{eqnarray}
These relations mean that the supersymmetric Wilson loop operator
is invariant if we perform supersymmetry transformations
(\ref{susy2}) simultaneously with the 
supersymmetry transformations for $(\lambda, k)$.
By expanding $\omega(\lambda,k)$ in terms of 
an appropriate basis of wave functions for $\lambda$ as
\be
\omega(\lambda,k) = \sum_A f_A(\xi) \ V_A(A_\mu, \psi; \ k),
\ee
we can define supersymmetry transformations for the wave functions
by
\bea
\delta^{(1)} f(\lambda,k) &=& \epsilon {\partial \over \partial\lambda}
f(\lambda,k), 
\label{susy5-1}\\
\delta^{(2)} f(\lambda,k) &=& (\bar\epsilon \kslash \lambda) f(\lambda,k).
\label{susy5-2}
\eea
These transformations are the same as (\ref{susy4}) except that these 
two supersymmetries are interchanged. As we explain later, the
interchanging can be realized by a charge conjugation operation.

The Majorana-Weyl fermion $\lambda$ contains 16 degrees of freedom
and there are $2^{16}$ independent wave functions for $\lambda$.
To reduce the number, we impose massless condition for the momenta $k$.
Then since $\kslash \lambda$ has only 8 independent degrees of
freedom the supersymmetry can generate only $2^8=256$
independent wave functions for $\lambda$.
They form a massless type IIB supergravity multiplet
containing a complex dilaton $\Phi$, 
a complex dilatino $\tilde\Phi$, a complex antisymmetric tensor 
$B_{\mu\nu}$, a complex gravitino $\Psi_\mu$, a real graviton 
$h_{\mu\nu}$ and a real 4-rank antisymmetric tensor $A_{\mu\nu\rho\sigma}$.

We now define a charge conjugation operation on the massless wave functions 
$f(\lambda, k)$. The charge conjugation is an operation to interchange 
a wave function with $p(\le 8)$ $\lambda$'s and that with $(8-p)$ $\lambda$'s.
It is defined by 
\be
(\hat{C} f)(\zeta, k) =
f^c(\zeta, k) \equiv \int [d\lambda] ~~\exp f(\lambda, k),
\label{chargeconjugation}
\ee
where the integration of $\lambda$ is performed with respect to 
eight $\lambda$'s included in $\kslash\lambda$.
The integral measure is normalized so that $\hat{C}^2=1$.
Acting $\hat{C}^2=1$ on a wave function, we get
\bea
 (\hat{C}^2 f)(\lambda', k) &=& \int [d \zeta][d \lambda]
e^{\bar\zeta \ \kslash (\lambda-\lambda')} f(\lambda, k)
= \int [d \zeta][d \lambda]
e^{\bar\zeta \ \kslash \lambda} f(\lambda+\lambda', k) \nonumber \\
&=& \int [d \zeta][d \lambda] \frac{1}{8!} (\bar\zeta \kslash \lambda)^8
f(\lambda+\lambda').
\eea
If we take a special momentum $k^\mu=(E, 0 \cdots 0, E)$
and use the Gamma matrices given in appendix, we have
\be
\frac{1}{8!} (\bar\zeta \kslash \lambda)^8 = (2E)^8 
(\zeta_9\zeta_{10} \cdots \zeta_{16})(\lambda_9 \cdots \lambda_{16}),
\ee
and the normalization of the integration is given by 
\be
\int [d\zeta] ~~(2E)^4(\zeta_9 \cdots \zeta_{16}) = 1.
\ee
It is easy to show that 
supersymmetry transformations for the charge conjugated fields are interchanged 
between $\delta^{(1)}$ and $\delta^{(2)}$;
\bea
(\delta^{(1)} f)^c(\zeta,k) &=& (\bar\epsilon \kslash \zeta) f^c(\zeta) 
= \delta^{(2)}f^c(\zeta), \\
(\delta^{(2)} f)^c(\zeta,k) &=& \epsilon{\partial \over \partial\zeta}~f^c(\zeta) =
 \delta^{(1)}f^c(\zeta).
\eea

\section{Wave Functions for the IIB Supergravity Multiplet}
\setcounter{equation}{0}
In this section we derive wave functions $f(\lambda, k)$ for a massless
supergravity multiplet by using the transformations (\ref{susy5-1}) and
(\ref{susy5-2}). It can be seen that these wave functions satisfy 
the susy transformations of the IIB supergravity. 

\subsection{Dilaton $\Phi$ and dilatino $\tilde\Phi$}
We start with the simplest wave function which can be interpreted as 
a dilaton field $\Phi$ in the IIB supergravity multiplet;
\be
\Phi(\lambda,k) = 1 \label{dilaton}.
\ee

Dilatino wave function  $\tilde\Phi$ can be generated from 
the dilaton wave function $\Phi$ by supersymmetry $\delta^{(2)}$ as  
\be
\delta^{(2)}\Phi(\lambda,k) 
= \bar\epsilon \kslash \lambda \equiv \bar\epsilon \tilde\Phi(\lambda,k).
\label{tr:dilaton2}
\ee
Hence the dilatino wave function is given by 
\be
\tilde\Phi(\lambda,k) = \kslash \lambda. 
\ee
The dilatino wave function automatically satisfies 
the equation of motion
\be
\kslash \tilde\Phi =0,
\ee
because of the massless condition $k^2=0$.
Then we can show the supersymmetry transformation
between the dilaton and the dilatino;
\be
\delta^{(1)}\tilde\Phi = \kslash \epsilon = \Gamma^\mu\epsilon (-i \partial_\mu\Phi).
\label{tr:dilaton1}
\ee

\subsection{Antisymmetric tensor field $B_{\mu\nu}$}
The wave function of the next field, an antisymmetric tensor field contains
two $\lambda$'s and can be generated from the dilatino wave function 
by $\delta^{(2)}$ transformation as
\be
\delta^{(2)} \tilde\Phi(\lambda,k) =
 -\frac{1}{16}\Gamma^{\mu\nu\rho}\epsilon \ 
 k_\mu \left(k^\sigma \bar{\lambda}\Gamma_{\nu\rho\sigma}\lambda\right)
 \equiv -{i \over 24} \Gamma^{\mu\nu\rho} \epsilon H_{\mu\nu\rho}, 
\label{tr:B2}
\ee
We identify $H_{\mu\nu\rho}$ as the field strength of the antisymmetric
tensor $B_{\mu\nu}(\lambda, k)$, 
\be
H_{\mu\nu\rho} = i(k_\mu B_{\nu\rho} +k_\nu B_{\rho\mu}+k_\rho B_{\mu\nu}).
\ee
{}Then the wave function $B_{\mu\nu}$ is given by
\be
B_{\mu\nu}(\lambda,k) = - {1\over 2} b_{\mu\nu} + 
(k_\mu v_\nu - k_\nu v_\mu)\equiv- {1\over 2} b_{\mu\nu} + 
k_{[\mu} v_{\nu]}, 
\label{Bwavefunction}
\ee
where $v_\mu$ represents gauge degrees of freedom 
corresponding to the two form gauge field $B_{\mu\nu}$. 
Here we have defined an antisymmetric bilinear of $\lambda$ by
\be
b_{\mu\nu}(\lambda) \equiv k^\rho (\bar\lambda \Gamma_{\mu\nu\rho}\lambda).
\ee
They are the only independent bilinear forms constructed 
from 8 independent massless spinors (namely nonzero components
of $\kslash \lambda$) and there are ${}_8 C_2=28$ degrees of freedom.
This number can be understood as follows.
$b_{\mu\nu}$ satisfies two relations
\bea
&& k^\mu b_{\mu\nu} =0, \\
&& b_{\mu\nu} \Gamma^{\mu\nu} \lambda =0,
\label{relation}
\eea
and an independent number of each relation is 9 and 8. Hence 
the number of independent $b_{\mu\nu}$ is ${}_{10}C_2-9-8=28.$
The proof of the second relation (\ref{relation}) is given in the appendix.

For simplicity we fix the gauge degrees of freedom as $v_\mu=0$.
For the wave function (\ref{Bwavefunction}), the equation of motion 
for the antisymmetric tensor is satisfied,
\begin{equation}
 k^\mu H_{\mu\nu\rho}=0,
\end{equation}
because of $k^2=0$ and 
\be
k^\mu B_{\mu \nu} =0.
\ee
A variation under the other supersymmetry $\delta^{(1)}$ 
of the wave function $B_{\mu \nu}(\lambda, k)$ is calculated as
\be
\delta^{(1)} B_{\mu\nu} = - \bar\epsilon \Gamma_{\mu\nu}\tilde\Phi.
\label{tr:B1}
\ee

\subsection{Gravitino $\Psi_\mu$}
A gravitino wave function contains three $\lambda$'s
and can be generated from $B_{\mu\nu}$ through $\delta^{(2)}$
supersymmetry transformation. It is defined through the susy transformation
\be
\delta^{(2)} B_{\mu\nu} = 2i(\bar\epsilon \Gamma_{[\mu}\Psi_{\nu]} 
+ k_{[\mu}\Lambda_{\nu]}).
\label{Btr2}
\ee
$\Lambda^\mu$ is a gauge transformation parameter. 
Since the left hand side of (\ref{Btr2}) becomes 
\be
\delta^{(2)} B_{\mu\nu} = (\bar\epsilon \kslash \lambda)
B_{\mu\nu}(\lambda,k)
= - {1\over 2}(\bar\epsilon \kslash \lambda) b_{\mu\nu} \label{Btr1},
\ee
we can identify the wave function
 \be
 \Psi_\mu(\lambda, k) 
 = - {i \over 24}(k_\rho \Gamma^{\nu\rho} \lambda) b_{\mu\nu},
 \label{gravitinowavefunction}
\ee
and 
the gauge transformation parameter 
\be
\Lambda_\mu(\lambda,k) = 
-{i \over 12} (\bar\epsilon \Gamma^\nu \lambda) b_{\mu\nu}.
\ee
The wave function (\ref{gravitinowavefunction}) 
automatically satisfies the equation of motion
\be
k_{\nu} \Gamma^{\mu \nu \rho} \Psi_{\rho} =0.
\ee
With the gauge choice in (\ref{gravitinowavefunction}), this equation of motion is
equivalent to
\be
\kslash \Psi_{\mu} =0,
\ee
because of 
\begin{equation}
 \Gamma^\mu \Psi_\mu = k^\mu \Psi_\mu = 0.
\end{equation}
The supersymmetry transformation $\delta^{(1)}$ for the gravitino wave function is
given by
\bea
\delta^{(1)} \Psi_\mu(\lambda,k) 
&=& -{i\over 24} \left[ (\Gamma^\nu\kslash\epsilon)b_{\mu\nu}
+ 2(\Gamma^\nu \kslash \lambda)(\bar\epsilon\Gamma_{\mu\nu\rho}\lambda)k^\rho  \right] 
\nonumber \\
&=& {1\over 24\cdot4} \left[ 9\Gamma^{\nu\rho}\epsilon H_{\mu\nu\rho} 
-\Gamma_{\mu\nu\rho\sigma}\epsilon H^{\nu\rho\sigma} \right] + \mathrm{(gauge~tr.)}.
\eea

\subsection{Graviton $h_{\mu\nu}$ and 4-rank antisymmetric tensor $A_{\mu\nu\rho\sigma}$}
In the wave functions containing four $\lambda$'s there are two fields, graviton $h_{\mu\nu}$
and 4-rank antisymmetric
tensor field $A_{\mu\nu\rho\sigma}$.
These wave functions can be read from the supersymmetry transformations of the gravitino field as
\be
\delta^{(2)} \Psi_\mu(\lambda,k) 
= {i \over 2} \Gamma^{\lambda\rho} k_\rho h_{\mu\lambda}\epsilon
+ {i \over 4 \cdot 5!} \Gamma^{\rho_1 \cdots \rho_5} \Gamma_\mu \epsilon F_{\rho_1 \cdots \rho_5}
+ \mathrm{(gauge~tr.)}. \label{Psitr}
\ee
Here the field strength $F_{\mu\nu\rho\sigma\tau}(\lambda, k)$  is defined by
\be
F_{\mu\nu\rho\sigma\tau} = i k_\mu A_{\nu\rho\sigma\tau} 
+ (\mbox{antisymmetrization})
= ik_{[\mu}A_{\nu\rho\sigma\tau]}.
\ee
Since the left hand side becomes
\bea
\delta^{(2)} \Psi_\mu(\lambda,k)
 &=& (\bar\epsilon \kslash \lambda) \Psi_\mu \\
 &=& 
 -{i \over 24} (\bar\epsilon \kslash \lambda)(\Gamma^\nu \kslash \lambda) b_{\mu\nu} 
 \nonumber \\
&=& {i \over 12 \cdot 16} \Gamma^{\nu\rho} \epsilon 
k_\rho {b_\mu}^{\sigma} b_{\sigma\nu} 
+{i \over 24 \cdot 16} \left[ {1 \over 5!}\Gamma^{\rho_1\rho_2\rho_3\rho_4} \epsilon 
k_{[\mu}b_{\rho_1\rho_2}b_{\rho_3\rho_4]}  - (\mathrm{gauge~tr.}) \right] ,
\nonumber \\
\eea
we have the graviton wave function  $h_{\mu\nu}$ as 
\be
h_{\mu \nu}(\lambda, k) = {1 \over 96} {b_\mu}^{\rho}b_{\rho\nu}.
\ee
Because of the identity $b_{\mu \nu} b^{\mu \nu}=0$, 
the graviton wave function is traceless. 
By using the self-duality of $F_{\mu\nu\rho\sigma\tau}$,
\bea
\Gamma^{\rho_1 \cdots \rho_5} \Gamma_\mu F_{\rho_1 \cdots \rho_5} 
&=& \Gamma^{\rho_1 \cdots \rho_5}_{~~~~~~\mu} F_{\rho_1 \cdots \rho_5}
+ 5\Gamma^{\rho_1 \cdots \rho_4} F_{\rho_1 \cdots \rho_4\mu} \\
&=& 10 \Gamma^{\rho_1 \cdots \rho_4} F_{\rho_1 \cdots \rho_4\mu}~,
\eea
we can also obtain the wave function for the field strength as
\be
F_{\rho_1 \cdots \rho_4\mu} = 
\frac{1}{32\cdot 4!}k_{[\mu}b_{\rho_1\rho_2}b_{\rho_3\rho_4]}.
\ee
and hence for the 4-rank antisymmetric tensor $A_{\rho_1 \cdots \rho_4}$ as
\be
A_{\rho_1 \cdots \rho_4}(\lambda,k) 
= -\frac{i}{32(4!)^2} b_{[\rho_1\rho_2}b_{\rho_3\rho_4]},
\ee
up to gauge transformations.
It can be checked directly that the field strength $F_{\mu\nu\rho\sigma\tau}$
is self-dual with this wave function.

Under the other susy transformation $\delta^{(1)}$, these wave functions
transform as follows,
\begin{eqnarray}
 \delta^{(1)}h_{\mu\nu} &=& 
  -\frac{i}{2}\bar{\epsilon}\Gamma_{(\mu}\Psi_{\nu)} 
  + \mbox{(gauge tr.)}, \\
 \delta^{(1)}A_{\mu\nu\rho\sigma} &=& 
  -\frac{1}{32\cdot 4!} \bar{\epsilon}\Gamma_{[\mu\nu\rho}\Psi_{\sigma]}
  + \mbox{(gauge tr.)},
\end{eqnarray}
where a round bracket for indices means symmetrization with a weight 1.

\subsection{Charge conjugation and the other wave functions}
The other wave functions in the massless supergravity multiplet can be 
similarly constructed by using the supersymmetry transformations.
In the following we instead make use of the charge conjugation operation 
(\ref{chargeconjugation}) to obtain the other wave functions.

First the charge conjugation of the dilaton field is given by
\be
\Phi^c(\zeta,k) = \int [d\lambda] ~~ \exp = (2E)^4(\zeta_9 \cdots \zeta_{16})
= \frac{1}{8\cdot8!}
b_\mu^{~\nu} b_\nu^{~\lambda} b_\lambda^{~\sigma} b_\sigma^{~\mu}(\zeta).
\ee
The determination of the coefficient is straightforward but not easy 
to obtain. We have determined the coefficient by using a computer and 
verified that it is consistent with the susy transformations of the 
wave functions.

The charge conjugated dilatino wave function becomes
\be
\tilde\Phi^c(\zeta,k) = \int [d\lambda] ~~ \exp \tilde\Phi(\lambda,k)
= \frac{1}{8!}k_\alpha \Gamma^{\mu \nu \alpha} \lambda
b_{\nu \rho} b^{\rho \sigma} b_{\sigma \mu}.
\ee
It also satisfies the same equation of motion as the dilatino field
\be
\kslash \tilde\Phi^c =0.
\ee
By taking the charge conjugation of the transformation (\ref{tr:dilaton2})
and (\ref{tr:dilaton1}), we have
\bea
&& \delta^{(1)} \Phi^c(\zeta,k) = \bar\epsilon \tilde\Phi^c(\zeta,k), \\
&& \delta^{(2)} \tilde\Phi^c(\zeta,k) = \Gamma^\mu\epsilon (-i \partial_\mu\Phi^c).
\eea

The wave function for the  charge conjugated antisymmetric tensor field 
is given by
\be
B_{\mu\nu}^c(\zeta,k) = \int [d\lambda] ~ \exp B_{\mu\nu}(\lambda, k)
= -\frac{1}{6!} b_{\mu\rho}b^{\rho\sigma}b_{\sigma\nu}.
\ee
{}From transformations (\ref{tr:B2}) and (\ref{tr:B1}), we have
supersymmetry transformations for the charge conjugated field as
\bea
\delta^{(1)} \tilde\Phi^c(\zeta,k) &=& 
- {i \over 24} \Gamma^{\mu\nu\rho} \epsilon (H_{\mu\nu\rho})^c, \\
\delta^{(2)} B_{\mu\nu}^c 
&=& -\bar\epsilon \Gamma_{\mu\nu} \tilde\Phi^c. 
\eea

Finally
the charge conjugated gravitino wave function becomes
\be
\Psi_\mu^c(\zeta,k) = \int [ d \lambda ] ~~ \exp \Psi_\mu(\lambda,k)
= -\frac{i}{4 \cdot 5!} k^\rho  \Gamma_{\rho \lambda}  \lambda b^{\lambda \sigma} b_{\mu \sigma},
\ee
and its supersymmetry transformation is given by
\bea
\delta^{(1)} B_{\mu\nu}^c 
&=& 2i(\bar\epsilon \Gamma_{[\mu}\Psi_{\nu]}^c + k_{[\mu}\Lambda_{\nu]}^c), \\
\delta^{(2)} \Psi_\mu^c (\zeta, k) 
&=& {1\over 24\cdot4} \left[ 9\Gamma^{\nu \rho}\epsilon (H_{\mu\nu\rho})^c
-\Gamma_{\mu\nu\rho\sigma}\epsilon (H^{\nu\rho\sigma})^c \right] 
+ \mathrm{(gauge~tr.)}.
\eea

Graviton and 4-rank antisymmetric tensor field are invariant under the charge conjugation:
\be
h_{\mu\nu}^c = h_{\mu\nu}~,~~A_{\mu\nu\rho\sigma}^c = A_{\mu\nu\rho\sigma}.
\ee
Therefore we have the charge conjugated supersymmetry transformation as
\bea
\delta^{(1)} \Psi_\mu^c
&=& {i \over 2} \Gamma^{\nu\rho} k_\rho h_{\mu\nu}\epsilon
+ {i \over 4 \cdot 5!} \Gamma^{\rho_1 \cdots \rho_5} \Gamma_\mu \epsilon F_{\rho_1 \cdots \rho_5}
+ \mathrm{(gauge~tr.)}, \\
 \delta^{(2)}h_{\mu\nu} &=& 
  -\frac{i}{2}\bar{\epsilon}\Gamma_{(\mu}\Psi_{\nu)}^c + \mathrm{(gauge~tr.)}, \\
 \delta^{(2)}A_{\mu\nu\rho\sigma} &=& 
  -\frac{1}{32\cdot 4!} \bar{\epsilon}\Gamma_{[\mu\nu\rho}\Psi_{\sigma]}^c+ \mathrm{(gauge~tr.)}.
\eea

\subsection{Wave functions and SUSY transformations}
We here summarize the wave functions for the massless multiplet and 
their supersymmetry transformations.
\begin{itemize}

\item Wave functions
\bea
\Phi(\lambda,k) &=& 1,  \nonumber \\
\tilde\Phi(\lambda,k) &=& \kslash \lambda, \nonumber \\
B_{\mu\nu}(\lambda,k) &=& - {1\over 2} b_{\mu\nu}(\lambda),
\nonumber \\
\Psi_\mu(\lambda,k) 
&=& - {i \over 24}(k_\sigma \Gamma^{\nu\sigma} \lambda) b_{\mu\nu}(\lambda), 
\nonumber \\
h_{\mu \nu}(\lambda,k) &=& {1 \over 96} b_\mu^{~\rho}b_{\rho \nu}(\lambda), 
\nonumber \\
A_{\mu\nu\rho\sigma}(\lambda,k)&=&-\frac{i}{32(4!)^2}
b_{[\mu\nu}b_{\rho\sigma]}(\lambda),
\nonumber \\
\Psi_\mu^c(\lambda, k) &=& 
-\frac{i}{4 \cdot 5!} k^\rho  \Gamma_{\rho \lambda}  
\lambda b^{\lambda \sigma} b_{\sigma\mu}(\lambda), \nonumber \\
B_{\mu\nu}^c(\lambda, k) 
&=& -\frac{1}{6!} b_{\mu\rho}b^{\rho\sigma}b_{\sigma\nu}(\lambda), 
\nonumber \\
\tilde\Phi^c(\lambda, k) &=& 
      \frac{1}{8!}k_\alpha \Gamma^{\mu \nu \alpha} \lambda 
      b_{\nu \rho} b^{\rho \sigma} b_{\sigma \mu}(\lambda), \nonumber \\
\Phi^c(\lambda, k) &=& \frac{1}{8 \cdot 8!}
b_\mu^{~ \nu} b_\nu^{~ \lambda} b_\lambda^{~ \sigma} b_\sigma^{~ \mu}(\lambda).
\eea

\item SUSY transformations
\bea
 \label{susy-transformation}
\delta \Phi &=& \bar\epsilon_2 \tilde\Phi, \nonumber \\
\delta \tilde\Phi &=& \kslash \epsilon_1 \Phi -{i \over 24} 
\Gamma^{\mu\nu\rho} \epsilon_2 H_{\mu\nu\rho}, \nonumber \\
\delta B_{\mu\nu} &=& - \bar\epsilon_1 \Gamma_{\mu\nu}\tilde\Phi 
+ 2i(\bar\epsilon_2 \Gamma_{[\mu}\Psi_{\nu]} + k_{[\mu}\Lambda_{\nu]}), 
\nonumber \\
\delta \Psi_\mu 
&=& {1\over 24\cdot4} \left[ 9\Gamma^{\nu \rho}\epsilon_1 H_{\mu\nu\rho} 
-\Gamma_{\mu\nu\rho\sigma}\epsilon_1 H^{\nu\rho\sigma} \right]
+{i \over 2} \Gamma^{\nu\rho} k_\rho h_{\mu \nu}\epsilon_2
\nonumber \\
&& + {i \over 4 \cdot 5!} \Gamma^{\rho_1 \cdots \rho_5} 
\Gamma_\mu \epsilon_2 F_{\rho_1 \cdots \rho_5} + k_\mu \xi, 
\nonumber\\
\delta h_{\mu\nu} &=& -\frac{i}{2} \bar\epsilon_1\Gamma_{(\mu}\Psi_{\nu)}
-\frac{i}{2} \bar\epsilon_2\Gamma_{(\mu}\Psi^c_{\nu)} 
+ k_{(\mu} \xi_{\nu)}, 
\nonumber \\
\delta A_{\mu\nu\rho\sigma} 
&=& 
-\frac{1}{(4!)^2} \bar\epsilon_1 \Gamma_{[\mu\nu\rho}\Psi_{\sigma]} 
-\frac{1}{(4!)^2} \bar\epsilon_2 \Gamma_{[\mu\nu\rho}\Psi^c_{\sigma]}
+ k_{[\mu} \xi_{\nu\rho\sigma]},
\nonumber \\
\delta \Psi_\mu^c
&=& {i \over 2} \Gamma^{\nu\rho} k_\rho h_{\mu \nu}\epsilon_1
+ {i \over 4 \cdot 5!} \Gamma^{\rho_1 \cdots \rho_5} 
\Gamma_\mu\epsilon_1 F_{\rho_1 \cdots \rho_5}
\nonumber \\
&& +{1\over 24\cdot4} \left[ 9\Gamma^{\nu \rho}\epsilon_2 H_{\mu\nu\rho}^c 
-{\Gamma_\mu}^{\nu\rho\sigma}\epsilon_2 H_{\nu\rho\sigma}^c \right]
+ k_\mu \xi^c, 
\nonumber \\
\delta B_{\mu\nu}^c 
&=& 2i(\bar\epsilon_1 \Gamma_{[\mu}\Psi_{\nu]}^c + k_{[\mu}\Lambda_{\nu]}^c)  
 -\bar\epsilon_2 \Gamma_{\mu\nu} \tilde\Phi^c,
\nonumber \\ 
\delta\tilde\Phi^c 
&=& - {i \over 24} \Gamma^{\mu\nu\rho} \epsilon_1 H_{\mu\nu\rho}^c 
+ \kslash\epsilon_2 \Phi^c, \nonumber \\
\delta \Phi^c &=& \bar\epsilon_1 \tilde\Phi^c,
\eea
\end{itemize}
where $\xi, \xi_\mu, \xi_{\mu\nu\rho}$ and $\Lambda_\mu$ are gauge parameters. 
This supersymmetry transformation is the same as that in \cite{Schwarz:1983wa}
up to  normalizations.

\section{Vertex Operators in IIB Matrix Model}
\setcounter{equation}{0}

In this section, we construct the vertex operators in IIB matrix model.
The construction can be done systematically by expanding 
the supersymmetric Wilson loop operator in terms of the wave functions 
$f_A(\lambda)$ constructed in the previous section. 

First we rewrite the Wilson loop operator (\ref{omega})  in terms of the supersymmetry transformations of 
$(i k\cdot A)$ as follows,
\be
\omega(\lambda,k)= e^{\bar\lambda Q_1}
~\mathrm{tr} e^{ik \cdot A}~
e^{- \bar\lambda Q_1} 
= \mathrm{tr} ~e^G,
\ee
where $G$ is given as a finite sum 
\bea
G&=&ik \cdot A + [\bar\lambda Q_1, ik \cdot A] 
+ {1 \over 2} [\bar\lambda Q_1,[\bar\lambda Q_1,ik \cdot A]] + \cdots
+{1 \over n!}[\bar\lambda Q_1, \cdots, [\bar\lambda Q_1,e^{ik \cdot A}] ]+ \cdots \nonumber \\
&=& \sum^{8}_{i=0} G_i. 
\eea
Note that the sum terminates at $i=8$ because there are only
8 independent $\lambda$'s for on-shell ($k^2=0$) Wilson loop operator.
Each term can be evaluated as follows;
\bea
G_0 &=& ik \cdot A, \\
G_1 &=& - (\bar\lambda \kslash \psi), \\
G_2 &=& {i \over 4} b^{\mu\nu}[A_\mu,A_\nu], \\
G_3 &=& -{1 \over 3!} b^{\mu\nu}[\bar\lambda\Gamma_\mu\psi,A_\nu], \\
G_4 &=& {1 \over 4!} 
\left\{
{i\over2}b^{\mu\nu}(\bar\lambda\Gamma_{\mu\rho\sigma}\lambda)
[[A^\rho,A^\sigma],A_\nu]
-ib^{\mu\nu}[\bar\lambda\Gamma_\mu\psi,\bar\lambda\Gamma_\nu\psi]
\right\}, \\
G_5 &=& -{1\over5!}
\left\{
b^{\mu\nu}(\bar\lambda\Gamma_{\mu\rho\sigma}\lambda)
[[\bar\lambda\Gamma^\rho\psi,A^\sigma],A_\nu]
+{3\over2}b^{\mu\nu}(\bar\lambda\Gamma_{\mu\rho\sigma}\lambda)
[[A^\rho,A^\sigma],\bar\lambda\Gamma_\nu\psi]
\right\}. \\
&\vdots& \nonumber
\eea
Note that $G_n$ contains $n$ $\lambda$'s.
In order to obtain a vertex operator of each field, we need to expand  
$\omega(\lambda,k)$ and collect all terms with the same number 
of $\lambda$ as
\bea
\omega(\lambda,k) 
&=& \mathrm{tr}~(e^{ik \cdot A + \sum^{8}_{i=1}G_i}) \nonumber \\
&=& \mathrm{Str}~e^{ik \cdot A} \biggl[ 
1+ G_1 + \left( {1\over 2}G_1\cdot G_1  + G_2 \right)
+ \left( \frac{(G_1^3)_\cdot}{3!} + G_1\cdot G_2 + G_3 \right) \nonumber \\
& & ~~~~~~~~ +\left(
 \frac{(G_1^4)_\cdot}{4!} + {1 \over 2}(G_1^2)_\cdot \cdot G_2 
 + {1\over 2}(G_2^2)_\cdot
  +G_1\cdot G_3 +G_4
 \right) \nonumber \\
& & ~~~~~~~~  +\bigg(
\frac{(G_1^5)_\cdot}{5!}
+{1\over3!}(G_1^3)_\cdot \cdot G_2
+{1\over2}G_1\cdot (G_2^2)_\cdot
+{1\over2}(G_1^2)_\cdot \cdot G_3 
+G_2\cdot G_3 
\nonumber \\
&& \hspace{20mm} 
+ G_1\cdot G_4
+ G_5
\bigg) 
\nonumber\\
& & ~~~~~~~~ + \cdots
\biggl] . 
\eea
Here "$\mathrm{Str}$" means a symmetrized trace which is defined by
\begin{eqnarray}
 \label{def-str}
 \mathrm{Str}~ e^{ik\cdot A} B_1\cdot B_2 \cdots B_n
  &=& 
  \int_0^1 dt_1 \int_{t_1}^1 dt_2 \cdots \int_{t_{n-2}}^1 dt_{n-1}
  \nonumber \\
 && 
  \times \mathrm{tr}~ e^{ik\cdot A t_1} B_1 e^{ik\cdot A(t_2-t_1)} B_2 
  \cdots 
  e^{ik\cdot A(t_{n-1}-t_{n-2})} B_{n-1} 
  e^{ik\cdot A(1-t_{n-1})} B_n
  \nonumber \\
 &&
  + \left(\mbox{ permutations of  $B_i$'s} \ (i=2, 3, \cdots, n)~\right).
\end{eqnarray}
The center-dot on the left hand side means that the operators $B_i$ 
are symmetrized. 
We denoted $\underbrace{G_k \cdot G_k \cdots G_k}_n$ as $(G_k^n)_\cdot$.
Various properties of the symmetrized trace is given in the appendix.
For notational simplicity we sometimes use $\mathrm{Str}$ with a single
operator like
$
 \mathrm{Str}~ (e^{i k \cdot A} B)
$
which is equivalent to an ordinary trace.
If we set $k=0$, the symmetrized trace becomes
\begin{equation}
\mathrm{Str}\left(B_1\cdot B_2 \cdots B_n\right)
= \frac{1}{n!}\sum_{perm.} 
\mathrm{tr}\left(B_{i_1}B_{i_2}\cdots B_{i_n}\right).
\end{equation}

\subsection{Dilaton $\Phi$ and dilatino $\tilde\Phi$}
Dilaton vertex operator  $V^{\Phi}$ is given by the leading order 
of $\lambda$, namely  a term without $\lambda$,
\be
V^\Phi =\mathrm{tr}~ e^{ik \cdot A}.
\ee
Dilatino vertex operator  $V^{\tilde\Phi}$ is read from the term with a single $\lambda.$
This is also easily obtained as
\be
\mathrm{tr}~e^{ik \cdot A}G_1
=
\mathrm{tr}~e^{ik \cdot A}(-\bar\lambda \kslash \psi) 
= (\mathrm{tr}~e^{ik \cdot A}\bar\psi) \cdot (\kslash \lambda),
\ee
\be
V^{\tilde\Phi} = \mathrm{tr}~e^{ik \cdot A}\bar\psi.
\ee
\subsection{Antisymmetric tensor field $B_{\mu \nu}$}
The vertex operator for the antisymmetric tensor $B_{\mu\nu}$ can be obtained
from the terms with two $\lambda$'s;
\bea
\mathrm{Str}~e^{ik \cdot A}\left( {1\over 2}G_1 \cdot G_1 +G_2 \right)
&=&
\mathrm{Str}~e^{ik \cdot A}\left( {1\over 2} (\bar\lambda\kslash\psi) \cdot (\bar\lambda\kslash\psi) 
+{i\over 4}b^{\mu\nu}[A_\mu,A_\nu] \right) 
\nonumber \\
&=& 
\mathrm{Str}~e^{ik \cdot A}\left( 
-{1\over32}k^\rho(\bar\psi \cdot \Gamma_{\mu\nu\rho}\psi) 
+{i \over 4}[A_\mu,A_\nu] 
\right)
\ b^{\mu\nu}.
\eea
Hence the vertex operator for the antisymmetric tensor field is given by
\bea
V^{B}_{\mu \nu} 
&=& \mathrm{Str}~e^{ik \cdot A}
\left( 
{1\over16}k^\rho(\bar\psi \cdot \Gamma_{\mu\nu\rho}\psi) 
-{i \over 2}[A_\mu,A_\nu] 
\right). 
\eea
This vertex operator satisfies 
\be
 k^{\mu} V^{B}_{\mu \nu}=0,
\ee
which assures the gauge invariance of the coupling with the wave function obtained in the
previous section,
$B^{\mu \nu}(\lambda) V^{B}_{\mu \nu}$.

\subsection{Gravitino $\Psi_\mu$}
The 3rd order terms give the gravitino $\Psi_\mu$ vertex operator as
\bea
&&
\mathrm{Str}~e^{ik\cdot A} 
\left(
{1\over 3!}G_1 \cdot G_1 \cdot G_1 +G_1 \cdot G_2+G_3
\right)
\nonumber \\
&& \hspace{10mm} = 
\mathrm{Str}~e^{ik\cdot A} 
\left(
-{1\over6}(\bar\lambda \kslash \psi)_{\cdot}^3 
- {i\over4}(\bar\lambda\kslash\psi) b^{\mu\nu} \cdot
[A_\mu,A_\nu]
-{1\over6}b^{\mu\nu}[A_\mu,\bar\lambda\Gamma_\nu\psi] 
\right).
\label{gravitino-1}
\eea
Here the following relation is useful,
\begin{equation}
 \label{fierz-gravitino}
 b_{\mu\nu}(\lambda)\left(\bar{\lambda} \kslash \psi\right)
  =\frac{1}{4}
  \left\{
   {b_\mu}^\sigma k^\rho (\bar{\lambda}\Gamma_{\sigma\nu\rho}\psi)
   -{b_\nu}^\sigma k^\rho (\bar{\lambda}\Gamma_{\sigma\mu\rho}\psi)
   -k_\mu {b_\nu}^\sigma (\bar{\lambda} \Gamma_\sigma \psi)
   +k_\nu {b_\mu}^\sigma (\bar{\lambda} \Gamma_\sigma \psi)
\right\}.
\end{equation}
Using this relation, the first term on the right hand side of
(\ref{gravitino-1}) becomes
\begin{equation}
 \left[
  -\frac{i}{12} \mathrm{Str}~e^{ik\cdot A}
  k^\rho \left(\bar{\psi} \cdot \Gamma_{\mu\nu\rho}\psi\right)
  \cdot \bar{\psi}\Gamma^\nu
\right]
 \Psi^\mu(\lambda),
\end{equation}
where $\Psi^\mu(\lambda)$ is the wave function of the gravitino
(\ref{gravitinowavefunction}).
Similarly the second term on the right hand side of (\ref{gravitino-1}) 
is rewritten as
\begin{equation}
 \mathrm{Str}~e^{ik\cdot A}
  \left[-\frac{i}{12}k^\rho {b_\mu}^\sigma
   \left(\bar{\psi}\Gamma_\nu \Gamma_{\rho\sigma}\lambda\right)
   \cdot [A_\mu, A_\nu]
 -\frac{1}{6}b^{\mu\nu}
 \left(\bar{\lambda}\Gamma_\nu \psi\right)
 \cdot [A_\mu, ik\cdot A]\right].
\end{equation} 
By using the relation (\ref{str-1}) in the appendix, it is easily
understood that the last term cancels the third term of
(\ref{gravitino-1}). 
Therefore the terms with three $\lambda$'s become 
\bea
&&
\mathrm{Str}~e^{ik\cdot A} 
\left(
{1\over 3!}(G_1^3)_\cdot+G_1 \cdot G_2+G_3
\right)
\nonumber \\
&=&
\mathrm{Str}~e^{ik\cdot A} 
\left(
-\frac{i}{12}k^\rho(\bar\psi \cdot \Gamma_{\mu\nu\rho} \psi) 
-2[A_\mu, A_\nu] 
\right)
\cdot \bar{\psi}\Gamma^\nu
\times \Psi^\mu(\lambda)
,
\eea
and thus we have the vertex operator for the gravitino 
\bea
V^{\Psi}_{\mu} 
&=& 
\mathrm{Str}~e^{ik\cdot A} 
\left(
-\frac{i}{12}k^\rho(\bar\psi \cdot \Gamma_{\mu\nu\rho} \psi) 
-2[A_\mu, A_\nu] 
\right)
\cdot \bar{\psi}\Gamma^\nu.
\eea
The second term is a matrix regularization of the supercurrent 
$\bar{J}_{\mu} = \{  X_{\mu}, X_\nu \}  \bar\psi \gamma^\nu$ 
associated with the supersymmetry  $\delta \psi =\epsilon'$ of the Schild action.
Here $\{ \ \  \}$ is Poisson bracket on the world sheet.

This gravitino vertex operator is shown to satisfy
\be
k^{\mu} V^{\Psi}_{\mu}=0.
\label{gra-vertex eq}
\ee
The first term of $V_\mu^\Psi$ trivially satisfies this relation 
and the second term is calculated as follows;
\bea
k^\mu
\biggl(
\mbox{the 2nd term of } V_\mu^\Psi
\biggl)
&=&
-2\mathrm{tr}\int_0^1dt~e^{ik \cdot At}
[k\cdot A, A_\mu] e^{ik\cdot A(1-t)} \bar\psi\Gamma^\mu
\nonumber \\
&=&
2i\mathrm{tr}~[e^{ik\cdot A},A_\mu]\bar\psi \Gamma^\mu
\nonumber\\
&=&
2i\mathrm{tr}~e^{ik\cdot A}[A_\mu,\bar\psi]\Gamma^\mu
\nonumber \\
&=&
0.
\label{gra-vertex cal}
\eea
In the last line, we used the equation of motion for the fermion, 
$\Gamma^\lambda[A_\lambda,\psi]=0$.
(\ref{gra-vertex eq}) assures the gauge invariance of the coupling with gravitino wave function
\be
V^{\Psi}_{\mu}\Psi^{\mu} . 
\ee

\subsection{Graviton and 4-rank antisymmetric tensor field}
The next terms with four $\lambda$'s  give the vertex operators for
the graviton $h_{\mu\nu}$ and the 4-rank antisymmetric tensor
$A_{\mu\nu\rho\sigma}$.
The calculation becomes more complicated and we need to use various 
identities involving fermions. Here we only write down the final results:
\bea
&&
\mathrm{Str}~e^{ik\cdot A} 
\left(
 {(G_1^4)_\cdot \over 4!} + {1 \over 2}G_1 \cdot G_1 \cdot G_2 
 + {1\over 2}G_2 \cdot G_2 +G_1 \cdot G_3 +G_4
 \right)
 \nonumber\\
&=&\mathrm{Str}~e^{ik\cdot A} 
\biggl(
{1\over4!}(\bar\lambda \kslash \psi)_{\cdot}^4
+
{i\over8}(\bar\lambda\kslash\psi)^2_{\cdot} b^{\mu\nu} \cdot [A_\mu,A_\nu] 
-
{1\over32}b^{\mu\nu}b^{\alpha\beta}[A_\mu,A_\nu]\cdot [A_\alpha,A_\beta]
\nonumber \\
&&~~~~~~~~~
+
{1\over6}(\bar\lambda \kslash \psi) b^{\mu\nu} \cdot 
[\bar\lambda \Gamma_\mu\psi,A_\nu]
-
{i\over24} b^{\mu\nu}
[\bar\lambda \Gamma_\mu\psi,\bar\lambda\Gamma_\nu\psi]
+
{i\over48}b_{\mu\nu}
(\bar\lambda \Gamma^{\alpha\beta\mu}\lambda)[[A_\alpha,A_\beta],A_\nu]
\biggr)
\nonumber \\
&=&
{1\over48}b^\mu_{~a} b^{a\nu}
\mathrm{Str}~e^{ik\cdot A}
\biggl\{
[A_\mu,A^\rho]\cdot [A_\nu,A_\rho] 
+ {1\over2}\bar\psi \cdot \Gamma_\mu[A_\nu,\psi] 
\nonumber \\
&&~~~~~~~~~~~~~~~~~~~~~~~
+{i\over4}k^\lambda (\bar\psi \cdot \Gamma_{\mu\lambda\sigma}\psi)\cdot [A^\sigma,A_\nu]
-{1\over8\cdot4!} k^\lambda k^\tau (\bar\psi \cdot \Gamma_{\mu\lambda}^{~~\sigma}\psi)
\cdot (\bar\psi \cdot \Gamma_{\nu\tau\sigma}\psi)
\biggl\}
\nonumber \\
&&+
{1\over3}\cdot \left(-{1\over32}\right)
(b^{\mu\nu}b^{\rho\sigma}+b^{\mu\rho}b^{\sigma\nu}+b^{\mu\sigma}b^{\nu\rho})
\mathrm{Str}~e^{ik\cdot A}
\nonumber \\
&&~~~~\times
\bigg\{
[A_\mu,A_\nu]\cdot[A_\rho,A_\sigma]
+C \bar{\psi}\cdot \Gamma_{\mu\nu\rho}[A_\sigma, \psi] 
-\frac{3i}{4}C k^\lambda(\bar\psi \cdot \Gamma_{\mu\nu\lambda}\psi)\cdot [A_\rho,A_\sigma]
  \nonumber \\
&&  \hspace{12mm}
-\frac{1}{8\cdot 4!} k^\lambda k^\tau (\bar\psi \cdot \Gamma_{\mu\nu\lambda}\psi)
\cdot (\bar\psi \cdot \Gamma_{\rho\sigma\tau}\psi)
\bigg\},
\eea
where $C$ is a numerical constant which  we could not determine in this approach of the 
calculation. But we can instead make use of other information of the block-block interaction briefly 
explained in the next subsection and  determine it to be $C=-1/3.$
Therefore we have the vertex operators for the graviton and the 4-rank 
antisymmetric tensor field respectively, 
%
%
%
%
%
\begin{eqnarray}
V^h_{\mu\nu}
&=&
2~\mathrm{Str}~e^{ik\cdot A}
\biggl\{
[A_\mu,A^\rho] \cdot [A_\nu,A_\rho] 
+ {1\over 4}\bar\psi \cdot \Gamma_{(\mu}[A_{\nu)},\psi]
-{i\over 8}k^\rho 
\bar\psi \cdot \Gamma_{\rho\sigma(\mu}\psi \cdot [A_{\nu)}, A^\sigma]
 \nonumber \\
&& \hspace{19mm}
-{1\over8\cdot4!} k^\lambda k^\tau 
(\bar\psi \cdot \Gamma_{\mu\lambda}^{~~~\sigma}\psi)
\cdot (\bar\psi \cdot \Gamma_{\nu\tau\sigma}\psi)
\biggl\}
,
\label{gravitonvertexop}
\end{eqnarray}
\begin{eqnarray}
 V^A_{\mu\nu\rho\sigma} 
  &=&
  -i~\mathrm{Str}~e^{ik\cdot A}
  \biggl\{
  F_{[\mu\nu} \cdot F_{\rho\sigma]} 
  +C\bar{\psi} \cdot \Gamma_{[\mu\nu\rho}[A_{\sigma]}, \psi]
  - \frac{3i}{4}C k^\lambda
  \bar\psi \cdot \Gamma_{\lambda[\mu\nu}\psi \cdot F_{\rho\sigma]}
  \nonumber \\
 && \hspace{35mm}
  -\frac{1}{8\cdot 4!} k^\lambda k^\tau 
  (\bar\psi \cdot \Gamma_{\lambda[\mu\nu}\psi)
  \cdot (\bar\psi \cdot \Gamma_{\rho\sigma]\tau}\psi)
  \biggr\},
  \label{4tensorvertexop}
\end{eqnarray}
where $F_{\mu\nu}=[A_\mu, A_\nu]$.
These vertex operators satisfy the conservation laws by similar calculations 
as (\ref{gra-vertex cal}),
\begin{equation}
\label{gravitoncons}
 k^\nu V^h_{\mu\nu}=0, \hspace{10mm}
 k^\sigma V^A_{\mu\nu\rho\sigma}=0,
\end{equation}
if we use the equations of motion (\ref{eom-boson}) and
(\ref{eom-fermion}). In the vertex operator of the graviton, 
while the fourth term trivially satisfies this equation,
the first three terms multiplied by $k_\mu$ are combined to become a term proportional 
to the equations of motion.
In the case of the 4-rank antisymmetric tensor, 
by multiplying $k_\mu$, the forth term trivially vanishes and so does 
the first term due to the Jacobi identity. 
The second and the third terms satisfy the conservation law because of 
properties of the symmetrized trace.

Thus the couplings with the graviton and the 4-rank antisymmetric tensor
wave functions 
\begin{equation}
 h^{\mu\nu}V^h_{\mu\nu}, \hspace{10mm}
  A^{\mu\nu\rho\sigma}V^A_{\mu\nu\rho\sigma},
\end{equation}
are respectively gauge invariant.

\subsection{Other vertex operators}
The other vertex operators are obtained from the terms containing more 
$\lambda$'s and the calculations of them become exponentially more difficult.
Therefore we do not proceed with this calculation here and give 
a part of the vertex operators by using other approaches.

The IIB matrix model can be regarded as a matrix regularization of 
the Schild type action for the IIB superstring. 
The supercurrent of the Schild action associated with the homogeneous 
supersymmetry (\ref{susy}) is given by
\begin{equation}
J_\mu^{(2)} =  
\{  X_\mu, X_\nu \} \{  X_\rho, X_\sigma \} 
\Gamma^{\rho \sigma} \Gamma^\nu \psi
-\frac{2i}{3}\left(\bar \psi \Gamma^\nu  \{X_\mu, \psi \}\right) 
\Gamma_\nu \psi.
\end{equation}
It is then expected that the vertex operator for the charge conjugation of 
the gravitino includes a term which is a matrix regularization of  
the above supercurrent of the Schild action.
Hence we have
\begin{equation}
  V_\mu^{\Psi^c} = 
   \mathrm{Str}~ e^{ik\cdot A} 
   \left( [A_\mu, A_\nu]\cdot [A_\rho, A_\sigma]\cdot 
    \Gamma^{\rho\sigma}\Gamma^{\nu} \psi
    +\frac{2}{3}\bar{\psi}\cdot \Gamma_{\nu}[A_\mu, \psi]\cdot
    \Gamma^{\nu} \psi
   \right).
\end{equation} 
This satisfies the relation $k^\mu V_\mu^{\Psi^c} = 0$ 
up to the equations of motion. Of course, the vertex operator will 
also contain other terms which include more fermions and momentum $k_\mu$. 

In the IIB matrix model, the interactions between supergravity modes 
can be obtained from the one-loop calculation 
by integrating out off-diagonal components of the matrices.
These interaction terms are interpreted as exchange of massless supergravity particles 
between vertex operators for the diagonal-blocks of the matrices.
Exchange of the graviton, dilaton and antisymmetric tensor field
is identified in \cite{IKKT} by calculation of one-loop effective action without
fermionic backgrounds. With fermionic backgrounds we can also identify 
exchange of the fermionic fields such as gravitinos and dilatinos.
Moreover we can also read off other terms of the bosonic vertex
operators containing even number of fermion fields such as the
second term of the  graviton vertex operator (\ref{gravitonvertexop}) or the
coefficient $C$ in the 4-th rank antisymmetric tensor field.
The one-loop effective action expanded with respect to the inverse
powers of the relative distance between two blocks was given in 
\cite{Taylor:1998tv}\cite{suyama}\cite{Kimura:2000ur};
\begin{eqnarray}
W^{(i,j)} &=&-
3 
S{\cal T}r^{(i,j)}
({\cal F}_{\mu\nu}{\cal F}_{\nu\sigma}{\cal F}_{\sigma\tau} {\cal F}_{\tau\mu}
-\frac{1}{4}{\cal F}_{\mu\nu} {\cal F}_{\mu\nu}{\cal F}_{\tau\sigma}{\cal F}_{\tau\sigma})
\frac{1}{(d^{(i)}-d^{(j)})^{8}} \cr
&&-
3 
S{\cal T}r^{(i,j)}
(\bar\Psi \Gamma^{\mu}\Gamma^{\nu}\Gamma^{\rho}
 {\cal F}_{\sigma\mu} {\cal F}_{\nu\rho}[{\cal A}_{\sigma}, \Psi])
\frac{1}{(d^{(i)}-d^{(j)})^{8}} \cr
&&+W_{\Psi^{4}}^{(i,j)} +O(\frac{1}{(d^{(i)}-d^{(j)})^{9}}) .
 \label{mmcalculation}
\end{eqnarray}
$W^{(i,j)}$ expresses the interaction between the $i$-th block 
and $j$-th block  and $(d^{(i)}-d^{(j)})$ is the distance 
between the center of mass coordinate of the 
$i$-th block and that of the $j$-th block. 
${\cal T}r$ is the trace of the adjoint operators and 
${\cal  F}$, ${\cal A}$ and $\Psi$ are adjoint operators which act as
${\cal O} S=[O,S].$
$W_{\Psi^{4}}$ denotes terms including four $\Psi$'s. 
The terms up to $O(r^{-7})$ cancel each other 
when backgrounds are restricted to satisfy the matrix model 
equations of motion.
{}From the above result  we can  identify some terms in the vertex operators.

In the case of the vertex operator for the charge conjugation of the 
antisymmetric tensor $V_{\mu\nu}^{B^c}$,
the leading term with the least number of fermion 
fields can be read from the calculations of the block-block interaction as,
\begin{equation}
  \mathrm{Str}~ e^{ik\cdot A}
  \left(
  [A_\mu, A_\rho]\cdot [A^\rho, A^\sigma]\cdot [A_\sigma, A_\nu]
  -\frac{1}{4}[A_\mu, A_\nu]\cdot [A^\rho, A^\sigma]\cdot [A_\sigma, A_\rho]
  \right).
\end{equation}
Requiring the current conservation, $k^\mu V_{\mu\nu}^{B^c}=0$, 
it can be understood that the vertex operator should include the following 
terms,
\begin{eqnarray}
  V_{\mu\nu}^{B^c} &=& \mathrm{Str}~ e^{ik\cdot A}
  \left(
  [A_\mu, A_\rho]\cdot [A^\rho, A^\sigma]\cdot [A_\sigma, A_\nu]
  -\frac{1}{4}[A_\mu, A_\nu]\cdot [A^\rho, A^\sigma]\cdot [A_\sigma, A_\rho]
  \right. \nonumber \\
  && \left.
  -\frac{1}{4}\bar{\psi}\cdot \Gamma_{(\mu}[A_{\rho)}, \psi]
  \cdot [A^\rho, A_\nu]
  +\frac{1}{4}\bar{\psi}\cdot \gamma_{(\nu}[A_{\rho)}, \psi]
  \cdot [A^\rho, A_\mu]
  \right. \nonumber \\
  && \left.
  +\frac{1}{16}\bar{\psi}\cdot\Gamma_{\rho\sigma[\mu}\psi
    \cdot \left[A_{\nu]}, [A^\rho, A^\sigma]\right]
  -\frac{i}{8}k_\lambda \bar{\psi}\cdot \Gamma^{\lambda\rho\sigma}\psi
  \cdot [A_\mu, A_\rho]\cdot [A_\nu, A_\sigma]
  \right).
\end{eqnarray}

For the charge conjugations of the dilaton,  
the leading terms of the vertex operators can be similarly read from block-block 
interactions as,
\begin{eqnarray}
  V^{\Phi^c} &=& 
  \mathrm{Str}~ e^{ik\cdot A}
  \bigg\{
   [A_\mu, A_\nu]\cdot [A^\nu, A^\rho]\cdot [A_\rho, A_\sigma]
   \cdot [A^\sigma, A^\mu]
    \nonumber \\
   && 
   -\frac{1}{4}[A_\mu, A_\nu] \cdot[A^\nu, A^\mu]
   \cdot [A_\rho, A_\sigma]\cdot [A^\sigma, A^\rho]
   \nonumber \\
   && 
   +  [A_\sigma,  A_\mu]   \cdot [A_\nu, A_\rho]  \cdot \bar\psi \Gamma^\mu \Gamma^{\nu \rho} \cdot  [A_\sigma, \psi]
   \bigg\}.
\end{eqnarray}
The charge conjugated dilatino vertex operator can be obtained from this
charge conjugated dilaton vertex operator by supersymmetry transformations.
The leading order term is proportional to 
\begin{eqnarray}
 V^{\tilde{\Phi}^c} &=&
  \mathrm{Str}~e^{ik\cdot A}
  \left\{
  \left(
  [A_\mu, A_\rho]\cdot [A^\rho, A^\sigma]\cdot [A_\sigma, A_\nu]
  -\frac{1}{4}[A_\mu, A_\nu]\cdot [A^\rho, A^\sigma]
  \cdot [A_\sigma, A_\rho]
  \right)
  \cdot \Gamma^{\mu\nu}\psi
  \right. \nonumber \\
  && \left.
  +\frac{1}{24}[A_\mu, A_\nu]\cdot [A_\rho, A_\sigma]
  \cdot [A_\lambda, A_\tau]
  \cdot \Gamma^{\mu\nu\rho\sigma\lambda\tau}\psi
  \right\}.
\end{eqnarray}
In order to obtain complete forms of the vertex operators, we need to 
accomplish the calculation which we performed in the previous section.
The calculation is very complicated and tough.
As we briefly explained above, we can instead determine the leading order terms of 
 the vertex operators  from the calculations of block-block interactions  with bosonic and 
fermionic backgrounds. 

\section{Conclusions and Discussions}
\setcounter{equation}{0}

In this paper, we have constructed a set of wave functions and vertex operators  
in the IIB matrix model by expanding the supersymmetric Wilson loop operator. 
They form a massless multiplet of the type IIB supergravity.
The vertex operators satisfy conservation laws, for instance 
eq.(\ref{gra-vertex eq}) or (\ref{gravitoncons}) 
by using equations of motion for $A^{\mu}$ and $\psi$. 

When we couple these vertex operators to background fields 
such as a graviton $h^{\mu \nu}$ field and integrate out the matrices, 
we can obtain effective action for the background fields.
Schematically it is written as
\begin{equation}
e^{-S_{eff}[h^{\mu \nu}]}= \int dA \ d\psi  \ e^{-S+\sum_k h^{\mu \nu} V^{h}_{\mu \nu}}.
\end{equation}
Since vertex operators satisfy the conservation laws, the effective action  
$S_{eff}[h^{\mu \nu}]$ has a gauge symmetry and it may be written as a sum of 
gauge invariant terms;
\begin{equation}
  \label{eff-action}
  S_{eff}[h^{\mu \nu}] = \int d^{10}x  
  \left(c_1(N) \sqrt{g} + c_2(N) \sqrt{g} R + \cdots \right).
\end{equation}
Here $c_i(N)$ are $N$ dependent coefficients. 
This is reminiscent of the induced gravity.  
Of course in order to show that the gravity theory indeed appears as above, 
we need to show that the graviton is formed as a bound state and  
calculate the coefficients as functions of the matrix size $N$. 
Both of them are very difficult to perform but we can instead make use 
of the large $N$ renormalization group as we will discuss below.

Here we discuss the origin of the conservation laws. 
We have used the supersymmetric Wilson loops and the supersymmetry 
transformations in order to obtain the vertex operators.
We did not use the explicit form of the action. 
Nonetheless the vertex operators satisfy the conservation laws 
by using the equations of motion derived from the action (\ref{IKKT}). 
This is due to the commutation relation of the supersymmetry generators 
(\ref{susycomm11}). Recall that the supersymmetric
Wilson loop is invariant under simultaneous supersymmetry transformations of
the matrices and wave functions as eqs.(\ref{susywilsoninv1})
and (\ref{susywilsoninv2}) only by using the equations of motion.
On the other hand, the supersymmetry transformations of the wave functions 
(\ref{susy-transformation}) contain  gauge transformations.
Because of it, the vertex operators satisfy conservation laws 
by using the equations of motion.
In this sense, the conservation laws for the vertex operators 
follow from the supersymmetries.
In string theories, conformal invariance guarantees the gauge invariance 
and the decoupling of unphysical modes from the S-matrix elements. 
It would be interesting to search for such a hidden symmetry in matrix models.

Another interesting issue is to obtain the equation of motion 
for the background field of the matrix models. 
In string theories, conformal invariance plays an important role in deriving
equation of motion for the background. 
In the matrix model, we expect that large $N$ 
renormalization group will play such a role. Matrix models are believed 
to describe string theories in the large $N$ limit. 
As we discussed in the introduction, we implicitly assume that 
there are background D(-1)'s other than the $N$ D(-1)'s and  
a modification of the configurations of the background D(-1)'s leads 
to a modification of the background field for the $N$ D(-1)'s. 
Hence stability of the background must be related to the stability 
of the background configurations under integrations
of the background D(-1)'s.  More concretely, we start from the matrix model for 
$(N+1) \times (N+1)$ hermitian matrices $A'_{\mu}$ with  a graviton coupling
\begin{equation}
 S_{IKKT}[A'_{\mu}] +  \int dk  \ h^{\mu \nu}(k) V^{h}_{\mu \nu}[A'_{\mu}],
\end{equation}
and integrate one D(-1) (which we call a mean field D(-1)).
Then we arrive at a matrix model action for $N \times N$ hermitian 
matrices $A_{\mu}$ with a modified graviton coupling
\begin{equation}
 S_{IKKT}[A_{\mu}] +  \int dk  \ h'^{\mu \nu}(k) V^{h}_{\mu \nu}[A_{\mu}],
\end{equation}
and we can obtain a renormalization group flow for the coupling constant
\begin{equation}
h^{\mu \nu}(k) \rightarrow h^{\mu \nu }(k) + \delta h^{\mu \nu}(k).
\end{equation}
Fixed points of this renormalization group flow will give the equations 
of motion for the background fields.
Though the calculation itself is very difficult, 
we can also in principle obtain renormalization group flow for 
the coefficients $c_i(N)$ of the effective action (\ref{eff-action}).
We want to investigate these issues in future publications.

\section*{Acknowledgements}
We would like to thank F. Sugino for his collaboration at the early stage
of this work and for his useful discussions and comments.
We would like to thank Drs. H. Aoki, K. Hamada, Y. Kitazawa, 
T. Suyama and A. Tsuchiya for useful discussions. 

\appendix
\section*{Appendix}
\setcounter{equation}{0}
\section{Majorana-Weyl representation}

We use the following Majorana-Weyl representation,
\begin{eqnarray}
 \Gamma_0&=&i \sigma_1\otimes1 \otimes 1 \otimes 1 \otimes 1, \nonumber \\ 
 \Gamma_1&=&i\epsilon\otimes\epsilon\otimes\epsilon\otimes\epsilon
  \otimes\epsilon, \nonumber \\
 \Gamma_2&=&i\epsilon\otimes\epsilon\otimes 1
  \otimes\sigma_1\otimes\epsilon, \nonumber \\ 
 \Gamma_3&=&i\epsilon\otimes\epsilon\otimes 1
  \otimes\sigma_3\otimes\epsilon, \nonumber \\ 
 \Gamma_4&=&i\epsilon\otimes\epsilon\otimes\sigma_1\otimes\epsilon\otimes 1,
  \nonumber \\ 
 \Gamma_5&=&i\epsilon\otimes\epsilon\otimes\sigma_3\otimes\epsilon\otimes 1, 
  \\ 
 \Gamma_6&=&i\epsilon\otimes\epsilon\otimes\epsilon\otimes 1
  \otimes\sigma_1, \nonumber \\ 
 \Gamma_7&=&i\epsilon\otimes\epsilon\otimes\epsilon\otimes 1
  \otimes\sigma_3, \nonumber \\ 
 \Gamma_8&=&i\epsilon\otimes\sigma_1\otimes 1 \otimes 1 \otimes 1, 
  \nonumber \\ 
 \Gamma_9&=&i\epsilon\otimes\sigma_3\otimes 1 \otimes 1 \otimes 1, 
  \nonumber \\ 
 \Gamma_{11}&=&\sigma_3\otimes 1 \otimes 1 \otimes 1 \otimes 1, \nonumber
\end{eqnarray}
where $\epsilon =i\sigma_2$.

\setcounter{equation}{0}
\section{Properties of gamma matrices}
\begin{itemize}
\item Metric
\be
g_{\mu\nu} = diag (-1,+1,\ldots,+1)\qquad(D=10)
\ee

\item Clifford algebra
\be
\{ \Gamma_\mu,\Gamma_\nu \} = 2g_{\mu\nu}
\ee
\be
\Gamma_{11} \equiv \Gamma_0 \Gamma_1 \cdots \Gamma_9~,~(\Gamma_{11})^2 = 1
\ee

\item Hermiticity
\be
(\Gamma_\mu)^\dagger = \Gamma^\mu = \Gamma_0 \Gamma_\mu \Gamma_0 
\ee
\be
(\Gamma_0)^\dagger = - \Gamma_0,~(\Gamma_i)^\dagger = \Gamma_i \quad (i=1,2,\ldots,9)
\ee
\be
(\Gamma_{11})^\dagger = \Gamma_{11}
\ee
Under our representations,
\be
(\Gamma_0)^\mathrm{T} = \Gamma_0~,~(\Gamma_i)^\mathrm{T} = -\Gamma_i~,
\ee
and
\be
\Gamma_0 \Gamma_\mu \Gamma_0 = -(\Gamma_\mu)^\mathrm{T}.
\ee

\item $\bar\psi \equiv \psi^\dagger \Gamma_0$
\be
(i \bar\psi \psi)^\ast = - i \psi^\dagger (\Gamma_0)^\dagger \psi = i \bar\psi \psi
\ee
\be
\left( \mathrm{tr} ~\bar\psi \Gamma_\mu [A^\mu,\psi]  \right) ^ \ast = \mathrm{tr} ~\bar\psi \Gamma_\mu[A^\mu,\psi]
\ee

\item Charge conjugation
\be
\psi^c = C \bar\psi^\mathrm{T} = \psi^\ast~~,~~C = \Gamma_0
\ee

\item Weyl spinor
\be
\psi = \Gamma_{11} \psi
\ee
\bea
\bar\psi_1 \Gamma_{\mu_1 \mu_2 \cdots \mu_n} \psi_2 &=& \psi^\dagger_1 \Gamma_0 \Gamma_{\mu_1 \mu_2 \cdots \mu_n} \Gamma_{11}\psi_2 \nonumber \\
&=&(-1)^{n+1} \bar\psi_1 \Gamma_{\mu_1 \mu_2 \cdots \mu_n} \psi_2
\eea
Therefore bilinear forms of spinors vanish unless $n$ is odd.
\item Majorana spinor
\be
\psi^c = \psi \quad \longrightarrow ~~ \psi = \psi^\ast
\ee
Under our representations,
\bea
\bar\psi_1 \Gamma_{\mu_1\mu_2\cdots\mu_n} \psi_2 &=&
- \psi^\mathrm{T}_2 (\Gamma_{\mu_1\mu_2\cdots\mu_n})^\mathrm{T}(\Gamma_0)^\mathrm{T}\psi^\ast_1
\nonumber \\
&=&-(-1)^{n(n-1) \over 2} \bar\psi_2 \Gamma_{\mu_1\mu_2\cdots\mu_n}\psi_1
\eea
When $\psi_1 = \psi_2$ is Majorana-Weyl spinor, therefore,  
bilinear forms of spinors vanish unless $n=3$ or 7.

\end{itemize}

\setcounter{equation}{0}
\section{Fierz identity}
The Fierz identity is given by \cite{Bergshoeff};
\be
(\bar\psi_1 M \psi_2)(\bar\psi_3 N \psi_4) = -{1\over32} \sum^5_{n=0} C_n 
(\bar\psi_1 \Gamma_{A_n} \psi_4)(\bar\psi_3 N \Gamma_{A_n} M \psi_2),
\ee
\be
C_0=2,~~~~C_1=2,~~~~C_2=-1,~~~~
C_3=-{1\over3},~~~~C_4={1\over12},~~~~C_5={1\over120}.
\ee
where $A_n$ is indexes for n-rank Gamma matrix. 
%
%

We here note some useful relations related to the Fierz identity.
We set
\bea
A &=& f^{\alpha\beta\gamma}(\bar\xi \Gamma_{\alpha\beta\gamma} \xi)\bar\xi, \\
B &=& f^{\alpha\beta\gamma}(\bar\xi \Gamma_{\nu\beta\gamma} \xi)
\bar\xi \Gamma_{\alpha}^{~\nu}, \\
C &=& f^{\alpha\beta\gamma}(\bar\xi \Gamma_{\mu\nu\alpha} \xi)
\bar\xi \Gamma^{\mu\nu}_{~~~\beta\gamma},
\eea
where $f^{\alpha\beta\gamma}$
is an arbitrary antisymmetric tensor.
Performing the Fierz transformation, we find 
\bea
A &=& -{1\over 32}(2A+6B-3C),  \\
B &=& -{1 \over 32}(14A+10B+3C).
\eea
{}From these relations we obtain 
\bea
(\bar\xi \Gamma_{\alpha\beta\gamma} \xi)\bar\xi \Gamma^{\alpha\beta}=0.
\eea

The following relation holds,
\be
2X +Y -Z =0,
\ee
where
\bea
X &=& f^{\mu\nu} k^\rho (\bar\epsilon \kslash \xi)(\bar\xi \Gamma_{\mu\nu\rho}\xi), \\
Y &=& f^{\mu\nu} k^\rho k_\nu (\bar\epsilon \Gamma^a \xi)
(\bar\xi \Gamma_{a\mu\rho}\xi), \\
Z &=& f^{\mu\nu} k^\rho k^\sigma (\bar\epsilon \Gamma^a_{~ \nu \sigma} \xi)
(\bar\xi \Gamma_{a \mu\rho} \xi).
\eea
Here $f^{\mu\nu}$ is an arbitrary antisymmetric tensor and $k^2 =0$.

We can derive the following identity from the Fierz transformation,
\bea
b_{\mu\nu}b_{\rho\sigma}
&=&
{1\over3}
(b_{\mu\nu}b_{\rho\sigma}+b_{\sigma\nu}b_{\mu\rho}-b_{\sigma\mu}b_{\nu\rho})
\nonumber \\
&+&
{1\over6}
(g_{\sigma\mu}b_{\nu}^{~\alpha}b_{\alpha\rho}-g_{\sigma\nu}b_{\mu}^{~\alpha}b_{\alpha\rho}
+g_{\rho\nu}b_{\mu}^{~\alpha}b_{\alpha\sigma}-g_{\rho\mu}b_{\nu}^{~\alpha}b_{\alpha\sigma})
\nonumber \\
&+&
{1\over6}
(k_\nu b_{\mu}^{~\alpha} - k_\mu b_{\nu}^{~\alpha})
(\bar\lambda \Gamma_{\rho\sigma\alpha}
\lambda)
\nonumber \\
&+&
{1\over6}
(k_\sigma b_{\rho}^{~\alpha} - k_\rho b_{\sigma}^{~\alpha})(\bar\lambda \Gamma_{\mu\nu\alpha}
\lambda),
\eea
where $b_{\mu\nu} = k^\rho(\bar\lambda \Gamma_{\mu\nu\rho}\lambda)$.

The following relations among the gamma matrices hold,
\bea
&&\Gamma^\mu \Gamma_{A_n} \Gamma_\mu = (-1)^n (10- 2n) \Gamma_{A_n},
\nonumber \\
&& \Gamma_{\alpha\beta\gamma}\Gamma_{\mu}\Gamma^{\alpha\beta\gamma}
    = 288 \Gamma_\mu, 
  \quad
  \Gamma_{\alpha\beta\gamma}\Gamma_{\mu\nu\rho}\Gamma^{\alpha\beta\gamma}
  = -48\Gamma_{\mu\nu\rho}, 
  \quad
  \Gamma_{\alpha\beta\gamma}\Gamma_{\mu\nu\rho\sigma\lambda}
  \Gamma^{\alpha\beta\gamma} = 0.
\eea

\setcounter{equation}{0}
\section{Symmetrized trace}
The symmetrized trace is defined in (\ref{def-str}).
In particular, explicit forms for two and three operators are written as
\begin{eqnarray}
 {\rm Str}\left(e^{ik\cdot A}B \cdot C\right)
  &=& {\rm tr} \int_0^1 dt \ e^{ik\cdot At}B e^{ik\cdot A(1-t)}C, \\
 {\rm Str}\left(e^{ik\cdot A}B \cdot C \cdot D\right) 
  &=& {\rm tr} \int_0^1 dt_1 \int_{t_1}^1 dt_2 
\ 
  e^{ik\cdot At_1}B e^{ik\cdot A(t_2-t_1)}C e^{ik\cdot A (1-t_2)} D 
  \nonumber \\ 
  && +  (C \longleftrightarrow D)
\end{eqnarray}
where all matrices are bosonic. The definitions for fermionic matrices 
can be obtained by replacing the bosonic matrices on the above equations 
with the fermionic matrices multiplied by Grassmann odd numbers.
The center-dot on the left hand side means that matrices are inserted at
different places.
We note useful equations related to the symmetrized trace,
\begin{eqnarray}
\label{str-1}
 {\rm Str}\left(e^{ik\cdot A}\left[ik\cdot A, \ A_\alpha\right]
	   \cdot \psi_\beta\right)
  &=& {\rm tr} \left[e^{ik\cdot A}, \ A_\alpha\right]\psi_\beta, \\
 {\rm Str}\left(e^{ik\cdot A}\bar{\psi}\cdot\Gamma_{\mu\nu\lambda}\psi
  \cdot\left[ik\cdot A, \ A_\rho\right]\right)
  &=& \label{str-2} 
  2~{\rm Str}\left(e^{ik\cdot A} \bar{\psi}\cdot\Gamma_{\mu\nu\lambda}
  \left[A_\rho, \ \psi\right]\right),
\end{eqnarray}
where the following relation is used,
\begin{equation}
 \label{str-3}
 \left[e^{ik\cdot A}, \ B\right]
  = \int_0^1 dt \ e^{ik\cdot At} \left[ik\cdot A, \ B\right]e^{ik\cdot A(1-t)}.
\end{equation}


\begin{thebibliography}{99}
\bibitem{IKKT}
N.~Ishibashi, H.~Kawai, Y.~Kitazawa and A.~Tsuchiya,
``A large-N reduced model as superstring,''
Nucl.\ Phys.\ B {\bf 498}, 467 (1997)
[arXiv:hep-th/9612115].

\bibitem{FKKT}
M.~Fukuma, H.~Kawai, Y.~Kitazawa and A.~Tsuchiya,
``String field theory from IIB matrix model,''
Nucl.\ Phys.\ B {\bf 510}, 158 (1998)
[arXiv:hep-th/9705128].


\bibitem{AIKKT}
H.~Aoki, S.~Iso, H.~Kawai, Y.~Kitazawa and T.~Tada,
``Space-time structures from IIB matrix model,''
Prog.\ Theor.\ Phys.\  {\bf 99}, 713 (1998)
[arXiv:hep-th/9802085].

\bibitem{Witten}
E.~Witten,
``Bound states of strings and p-branes,''
Nucl.\ Phys.\ B {\bf 460}, 335 (1996)
[arXiv:hep-th/9510135].


\bibitem{Yoneya}
T.~Yoneya,
``Supergravity, AdS/CFT correspondence, and matrix models,''
Prog.\ Theor.\ Phys.\ Suppl.\  {\bf 134}, 182 (1999)
[arXiv:hep-th/9902200].

\bibitem{Kitazawa:2002vh}
Y.~Kitazawa,
``Vertex operators in IIB matrix model,''
JHEP {\bf 0204}, 004 (2002)
[arXiv:hep-th/0201218].

\bibitem{Taylor:1998tv}
W.~I.~Taylor and M.~Van Raamsdonk,
``Supergravity currents and linearized interactions for matrix theory
configurations with fermionic backgrounds,''
JHEP {\bf 9904}, 013 (1999)
[arXiv:hep-th/9812239].


\bibitem{Dasgupta:2000df}
A.~Dasgupta, H.~Nicolai and J.~Plefka,
``Vertex operators for the supermembrane,''
JHEP {\bf 0005}, 007 (2000)
[arXiv:hep-th/0003280].

\bibitem{matrixstring}
R.~Schiappa,
``Matrix strings in weakly curved background fields,''
Nucl.\ Phys.\ B {\bf 608}, 3 (2001)
[arXiv:hep-th/0005145].

\bibitem{Hamada}
K.~J.~Hamada,
``Supersymmetric Wilson loops in IIB matrix model,''
Phys.\ Rev.\ D {\bf 56}, 7503 (1997)
[arXiv:hep-th/9706187].

\bibitem{Schwarz:1983wa}
J.~H.~Schwarz and P.~C.~West,
``Symmetries And Transformations Of Chiral N=2 D = 10 Supergravity,''
Phys.\ Lett.\ B {\bf 126}, 301 (1983).

\bibitem{suyama}
T. Suyama and A. Tsuchiya (unpublished), 
talk at JPS meeting 1999 and private communication

\bibitem{Kimura:2000ur}
Y.~Kimura and Y.~Kitazawa,
``Supercurrent interactions in noncommutative Yang-Mills and IIB matrix
model,''
Nucl.\ Phys.\ B {\bf 598}, 73 (2001)
[arXiv:hep-th/0011038].




\bibitem{Bergshoeff}
E.~Bergshoeff, M.~de Roo, B.~de Wit and P.~van Nieuwenhuizen,
``Ten-Dimensional Maxwell-Einstein Supergravity, Its Currents, And The Issue
Of Its Auxiliary Fields,''
Nucl.\ Phys.\ B {\bf 195}, 97 (1982).




\end{thebibliography}
\end{document}